\documentclass[lettersize,journal]{IEEEtran}
\usepackage{amsmath,amsfonts}
\usepackage{algorithmic}
\usepackage{algorithm}
\usepackage{array}
\usepackage[caption=false,font=normalsize,labelfont=sf,textfont=sf]{subfig}
\usepackage{textcomp}
\usepackage{stfloats}
\usepackage{url}
\usepackage{verbatim}
\usepackage{graphicx}
\usepackage{cite}
\usepackage{multirow}
\usepackage{booktabs}
\usepackage{bm}
\usepackage{color}
\usepackage{makecell}
\begin{document}

\title{An innovation-based cycle-slip, multipath estimation, detection and mitigation method for tightly coupled GNSS/INS/Vision navigation in urban areas}

\author{Bo Xu, Shoujian Zhang, Jingrong Wang, Jiancheng Li
\thanks{Bo Xu and  Shoujian Zhang are with School of Geodesy and Geomatics, Wuhan University, Wuhan 430079, China; Corresponding author: Shoujian Zhang, Email: shjzhang@sgg.whu.edu.cn
}
\thanks{Jingrong Wang is with the GNSS research center, Wuhan University, Wuhan 430079, China;}
\thanks{Jiancheng Li is with the School of Geodesy and Geomatics, Hubei Luojia Laboratory, Wuhan University, Wuhan 430079, China;}}



\maketitle

\begin{abstract}
Precise, consistent, and reliable positioning is crucial for a multitude of uses. In order to achieve high precision global positioning services, multi-sensor fusion techniques, such as the Global Navigation Satellite System (GNSS)/Inertial Navigation System (INS)/Vision integration system,
combine the strengths of various sensors. This technique is essential for localization in complex environments and has been widely used in the mass market. 
However, frequent signal deterioration and blocking 
in urban environments exacerbates the degradation of GNSS positioning and negatively impacts the performance of the multi-sensor integration system. For GNSS pseudorange and carrier phase observation data in the urban environment, we offer an innovation-based cycle slip/multipath estimation, detection, and mitigation (I-EDM) method to reduce the influence of multipath effects and cycle slips on location induced by obstruction in urban settings. The method obtains the innovations of GNSS observations with the cluster analysis method. Then the innovations are used to detect the cycle slips and multipath. Compared with the residual-based method, the innovation-based method avoids the residual overfitting caused by the least square method, resulting in better detection of outliers within the GNSS observations. The vehicle tests carried out in urban settings verify the proposed
approach. Experimental results indicate that the accuracy of 0.23m, 0.11m, and 0.31m in the east, north and up components can be achieved by the GNSS/INS/Vision tightly coupled system with the I-EDM method, which has a maximum of 21.6\% improvement when compared with the residual-based EDM (R-EDM) method.  
\end{abstract}



\section{Introduction}
\IEEEPARstart{A}{utonomous} driving has captured significant attention in both research and practical applications, where GNSS serves as the primary positioning technology to provide continuous, accurate, and stable positioning results \cite{joubert2020developments, xiong2021g, dasgupta2022sensor,10587194 }. Despite the worldwide prevalence and extensive utility of GNSS, its signals are susceptible to disruptions such as multipath effects and signal loss in complex urban settings, resulting in decreased positioning accuracy \cite{liao2021enhancing}. Consequently, the utilization of multi-sensors provides a more comprehensive and robust positioning capability, which becomes the optimal choice for pose estimation in urban scenarios \cite{10285614, 10385063, 10334484, dou2023novel, jiang2021r2}.

Integrating different sensors, such as INS and cameras, with GNSS can fully utilize the locally accurate characteristics of INS/Visual Inertial Odometry (VIO) and the global drift-free characteristics of GNSS. Fusing GNSS/INS has been widely studied. Significant improvements in accuracy, continuity, and reliability of localization are carefully analyzed and evaluated  \cite{angrisano2010gnss, dong2020tightly, chen2020estimate, gao2016real, liu2016tight}. However, in an integrated system where GNSS serves as the cornerstone for providing global positioning, the positioning accuracy is compromised or even worsens due to the rapid error accumulation of the microelectromechanical system inertial measurement unit (MEMS-IMU) when GNSS observations are affected by multipath effects or lost tracking  \cite{li2021improving}.

To mitigate the drifts of MEMS-IMU, the low-cost visual camera is applied in GNSS/INS integration system to offer redundant measurements. The multi-sensor fusion extended Kalman filter (MSF-EKF) framework proposed by Lynen et al. \cite{lynen2013robust} allows for the seamless loose sensor-feed integration of GNSS, INS, and visual sensors. A semi-tightly coupled framework of multi-GNSS/INS/Vision based on graph optimization is proposed by Li et al. \cite{li2021semi}. The method produces consistent and accurate global positioning outputs and operates well in GNSS-challenged environments. A tightly coupled GNSS/INS/Vision system with open-source code is proposed by Cao et al. \cite{cao2022gvins}; However, the method only incorporates Doppler shift and code pseudorange measurements. To enhance the navigation accuracy during GNSS outages, Liao et al. \cite{liao2021enhancing} propose a tightly coupled framework to make full use of measurements from real-time kinematic (RTK)/INS/Vision. The method improves the accuracy and robustness of the system under conditions where GNSS is unavailable.  However, these methods regard GNSS observations as equal weight, without considering the impact of outliers on the system, thereby failing to make the most of the GNSS observations.

Effective GNSS measurement quality control is key to ensuring that the estimator is not affected by outliers in urban situations. A practical quality control technique is to assign weights according to the quality of the signals.  The common GNSS weighting strategies include: signal to noise ratio (SNR) model \cite{hartinger1999variances}, satellite elevation angle model \cite{gao2011improved}, SNR and satellite elevation angle hybrid model \cite{herrera2016gogps}  et al. However, computing the appropriate weights for GNSS observations in urban situations is challenging. Leveraging sky plots to divide the satellites into line-of-sight (LOS) and non-line-of-sight (NLOS), and decreasing the weights of NLOS can increase the positioning accuracy in urban settings \cite{wen2019tightly, wang2024skygvio}. Nevertheless, this approach is limited to adjusting weights at the satellite levels, which means it can not modify the weights corresponding to different frequencies from the same satellite. An additional technique for measurement quality control is fault detection and exclusion (FDE) \cite{zhu2018extended, jurado2020residual, sun2021new, jiang2022effective}. A multiple-fault GNSS FDE technique is proposed by Sun et al.    \cite{sun2021new}  for an integrated GNSS/IMU system. A parallel GNSS FDE technique for tightly coupled GNSS/INS/Vision integration through factor graph optimization is presented by Jiang et al.  \cite{jiang2022effective}. However, they only perform quality control on the pseudorange observations and do not consider the carrier phase observations.

Even though significant progress has been made in the area of positioning in challenging situations, the GNSS quality control still requires extensive attention. In our earlier work, a unified cycle slip, multipath estimation, detection, and mitigation (EDM) method is proposed \cite{xu2023unified}, in which we utilize the clustering method to separate the cycle slips and multipath from the carrier phase observations aided by the predicted VIO positioning. However, this methodology entails repeatedly estimating the parameters using the least square method, followed by cycle slips detection and multipath modeling with the observation residuals. Whereas, the residuals are inevitably absorbed into the estimated parameters (e.g. positioning and receiver clock) with the least square method, leading to inaccurate outlier culling and cycle slip detection. Meanwhile, the method relies on the predicted VIO position to preprocess the GNSS observations. In the case of GNSS blocking frequently, setting the uncertainty of the predicted position becomes challenging. This contribution proposes an innovation-based EDM method. The innovations are applied to identify the cycle slips and multipath in the GNSS observations. To obtain the innovations, the inter-frequency bias (IFB) is estimated with the clustering method, and the receiver clock is obtained using the satellite with the highest elevation angle. Furthermore, considering the persistent impact of multipath on ambiguities, we mark the ambiguity as the cycle clip when the accumulated multipath exceeds a certain threshold to mitigate the influence of multipath effect on localization. Finally, we conduct a detailed analysis of the effectiveness of the innovation-based EDM method and residual-based EDM method. 

Our main contributions are summarized as follows: 
\begin{itemize}
\item We proposed an innovation-based EDM method for the pseudorange and carrier phase observations in the tightly coupled GNSS/MEMS/Vision system.  
\item We compare and analyze the residual-based (R-EDM) and innovation-based (I-EDM) methods in terms of cycle slip detection, multipath estimation, positioning accuracy, and computational efficiency in the tightly coupled multi-sensor fusion system. 
\item Extensive road vehicular experimental evaluations are conducted in the urban area to evaluate the performance of our method. The experimental results demonstrate the superior performance of our GNSS observation preprocessing method in complex urban settings.
\end{itemize}

The introduction is followed by a description of the RTK/MEMS-IMU/Vision integration approach and our proposed innovation-based EDM method. Then, the detailed description of the vehicle-borne experimental setups and processing strategies are described in detail. The experimental outcomes of different GNSS observation weighting schemes in typical urban settings are examined, and the effectiveness of the innovation-based EDM and residual-based EDM approaches is contrasted. The conclusions are given in the end.


\section{Methods}
This section begins with an overview of the tightly coupled RTK/MEMS/Vision system. The error models of all the associated senors are next presented, followed by the time update and measurement update model in the tightly coupled filter. Lastly, we provide the specifics of our proposed innovation-based EDM algorithm.

\begin{figure}[t]
  \centering
  \vspace{-1mm}
  \includegraphics[width=0.48\textwidth]{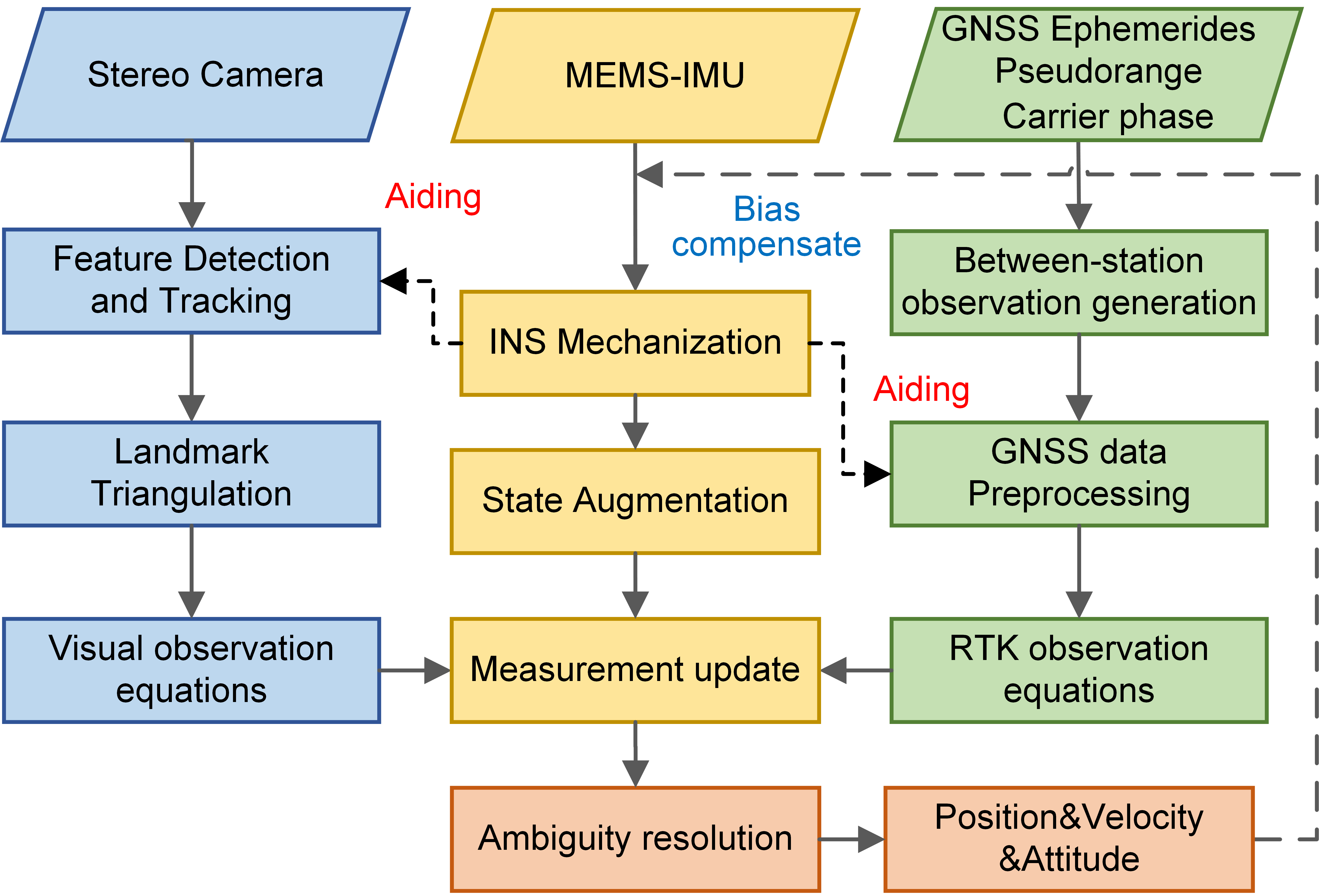}
  \vspace{-3mm}
  \caption{Implementation of RTK/MEMS/Vision tightly coupled system}
  \label{fig:framework}
  \vspace{-4mm}
\end{figure}
\subsection{System Overview}
Fig.\ref{fig:framework} shows the system architecture of the tightly coupled RTK/MEMS/Vision system. Based on the Multi-State Constraint Kalman Filter (MSCKF), the pseudorange and carrier phase observations from GNSS, the raw data of MEMS-IMU, and the images from the stereo camera are fused. The INS mechanization is used to carry out the state propagation once the system has been initialized. The predicted state variables will also help with feature matching and tracking in the image processing, as well as the preprocessing of outlier culling and cycle slip detection in GNSS processing. The state variables and associated covariance in the estimator will be augmented upon receiving new visual or GNSS observations. Once the new observations are deemed to be available, the corresponding state variables and the covariance will be updated in the measurement update. After completing the filter, the state variables are ultimately fixed through ambiguity resolution.

\subsection{INS Error Model}
The vehicle platform's biased noisy angular velocity and linear acceleration make up the measurements of the IMU. Because of the low-cost IMU measurement noise, the Coriolis and centrifugal forces caused by Earth's rotation are disregarded in IMU's formulation. Therefore, the kinematic model of the IMU's error state can be written as  \cite{sun2018robust}:
\begin{equation}
\begin{aligned}
    \delta \dot{\mathbf{p}}^n &= \delta \mathbf{v}^n \\
    \delta \dot{\mathbf{v}}^n &= -\mathbf{R}^n_b\left(\Tilde{\mathbf{a}}-\mathbf{b}_a\right)^{\wedge}
    \delta \bm{\theta}-\mathbf{R}^n_b\delta\mathbf{b}_a-\mathbf{R}_b^n \mathbf{n}_a \\
    \delta \dot{\bm{\theta}} &= -\left(\bm{\Tilde{\omega} - \mathbf{b}}_w\right)^{\wedge}\delta\bm{\theta}-\delta\mathbf{b}_w-\mathbf{n}_w \\
    \delta \dot{\mathbf{b}}_w &= \mathbf{n}_{b_w} \\
    \delta \dot{\mathbf{b}}_a &= \mathbf{n}_{b_a} \\
    \end{aligned}
\end{equation}
where $b$ and $n$ are the IMU body frame and navigation frame (East-North-Up frame), respectively. $\delta\dot{\mathbf{p}}^n$, $\delta\dot{\mathbf{v}}^n$, $\delta \dot{\bm{\theta}}$, $ \delta \dot{\mathbf{b}}_w$  and $\delta \dot{\mathbf{b}}_a$ represent the derivative of the position, velocity, attitude, gyroscope bias and accelerometer bias error in the navigation frame, respectively. $\mathbf{R}^n_b$ indicates the rotation from the body frame to the navigation frame. $\Tilde{\mathbf{a}}$ and $\Tilde{\bm{\omega}}$ are the acceleration and angular velocity measurement, respectively. The vector  $\mathbf{n}_a$ and $\mathbf{n}_w$ represent the Gaussian noise of accelerometer and gyroscope measurement, while $\mathbf{n}_{b_a}$ and $\mathbf{n}_{b_w}$ are the random walk rate of the accelerometer and gyroscope measurement biases. $\Tilde{\mathbf{a}}^{\wedge}$ is the skew symmetric matrix of $\Tilde{\mathbf{a}}$. Therefore, the INS error state vector has the following writing:

\begin{equation}
    \delta \mathbf{x}_{ins} = \left[\delta \bm{\theta} \quad \delta \mathbf{v}^n \quad \delta\mathbf{p}^n \quad \delta \mathbf{b}_w \quad \delta\mathbf{b}_a\right]^{\top}
\end{equation}

\subsection{Visual Observation Model}
Taking into account that the stereo camera observes a visual point feature $f^j$, the related visual observation measurement $\mathbf{z}_i^j$ can be written as follows:

\begin{equation}
    \mathbf{z}_i^j= \left(\begin{matrix}
    u_{i,1}^j \\ v_{i,1}^j \\u_{i,2}^j \\ v_{i,2}^j 
    \end{matrix}\right) = \left(\begin{matrix}
   \frac{1}{z_j^{C_{i, 1}}} \mathbf{I}_{2 \times 2} & \mathbf{0}_{2 \times 2} \\ \mathbf{0}_{2 \times 2} & \frac{1}{z_j^{C_{i, 2}}} 
    \end{matrix}\right) \left(\begin{matrix}
    x_j^{C_{i, 1}}\\ y_j^{C_{i, 1}} \\x_j^{C_{i, 2}} \\ y_j^{C_{i, 2}} 
    \end{matrix}\right) + \mathbf{n}^j_i
\end{equation}
where $\left[u_{i, k}^j, v_{i, k}^j\right]^{\top}, k \in \{1,2\}$ indicate the feature observations of the left and right cameras on their normalized projective plane. $\mathbf{n}^j_i$ indicates the visual measurement noise. $\left[x_j^{C_{i, n}},  y_j^{C_{i, n}}, z_j^{C_{i, n}}\right]^{\top}, n \in \{1,2\}$ are the visual landmarks in the camera frame, which can be calculated as follows:

\begin{equation}
\begin{aligned}
\left(\begin{matrix}
   x_j^{C_{i, 1}} \\y_j^{C_{i, 1}} \\z_j^{C_{i, 1}}\end{matrix}\right) &= \left(\mathbf{R}^n_{C_{i,1}} \right)^{\top} \left(\mathbf{p}^n_j - \mathbf{p}^n_{C_{i, 1}} \right) \\
\left(\begin{matrix}
   x_j^{C_{i, 2}} \\y_j^{C_{i, 2}} \\z_j^{C_{i, 2}}\end{matrix}\right) &= \mathbf{R}^{C_{i,2}}_{C_{i,1}} \left(\mathbf{p}^{C_{i,1}}_j - \mathbf{p}^{C_{i,1}}_{C_{i, 2}} \right)
   \end{aligned}
\end{equation}
where $\mathbf{R}^{C_{i,2}}_{C_{i,1}}$ and $\mathbf{R}^n_{C_{i,1}}$    denote the rotation matrix that goes from the left camera frame to the right camera frame and navigation frame, respectively. The location of the left camera with regard to the navigation frame is $\mathbf{p}^n_{C_{i, 1}}$. $\mathbf{p}^{C_{i,2}}_{C_{i, 1}}$ is the position of the left camera frame with respect to the right camera frame. $\mathbf{p}_j^n$ and $\mathbf{p}_j^{C_{i,1}}$ represent the visual landmarks' position in the navigation frame and left camera frame, respectively.

In order to construct the visual reprojection residuals between relative camera poses, we use the algorithm that is proposed by \cite{sun2018robust}. The following is the description of the visual state vector: 

\begin{equation}
    \delta \mathbf{x}_{vis} = \left[\delta \bm{\theta}^n_{C_1} \ \delta \bm{p}^n_{C_1} \  \delta \bm{\theta}^n_{C_2} \ \delta \bm{p}^n_{C_2} \ \cdots \ \delta \bm{\theta}^n_{C_k} \  \delta \bm{p}^n_{C_k}\right]^{\top}
\end{equation}
where $\delta\bm{\theta}^n_{C_i}$ and $\delta\bm{p}^n_{C_i}$ represent the error state of the left camera's rotation and position at various time stamps, respectively. $k$ indicates how many camera poses there are in the sliding window overall. 

The following is the expression for the projection residual of the visual measurement:

\begin{equation}
    \mathbf{r}_{vis} = \mathbf{z}_{vis} - \hat{\mathbf{z}}_{vis} = \mathbf{H}_{vis} \mathbf{x}_{vis} + \mathbf{n}_{vis}
\end{equation}
where $\mathbf{z}_{vis}$ and $\hat{\mathbf{z}}_{vis}$ are the observation and reprojection visual measurements, respectively, and $\mathbf{H}_{vis}$ is the Jacobian of the relevant camera states, which are provided in \cite{sun2018robust}.
\subsection{Between-station Single-difference Multi-GNSS Observation Model}
From the undifferenced pseudorange and carrier phase models, we will derive the between-station single-difference GNSS model.  The undifferenced pseudorange $P^s_{r,i}$ and carrier phase $L_{r,i}^s$ model is first given as follows:

\begin{equation}
    \begin{aligned}
    P^s_{r,i}&= \bm{\rho}+c\left(t_r-t^s\right)+T_r^s+I_{r,i}^s+e_{r,i}^s,\sigma^2_{P^s_{r,i}} \\  
     L_{r,i}^s&=\bm{\rho}+c(t_r-t^s)+T_r^s-I_{r,i}^s+ \lambda_i \cdot N_{r,i}^s+\epsilon_{r,i}^s, \sigma^2_{L_{r,i}^s}
    \end{aligned} \label{equ:gnss observation}
\end{equation}
where the satellite and receiver are denoted by $s$ and $r$, respectively. The carrier frequency is $i=1,2,3$. The speed of light is $c$. The pseudorange and carrier phase observations are denoted by $P_{r,i}^s$ and $L^s_{r,i}$. The distance between the receiver and the satellite is  $\bm{\rho}$. The receiver clock and satellite clock are $t_r$ and $t^s$. The tropospheric and ionospheric delays on the $i$ frequency are $T^s_r$ and  $I^s_{r,i}$.  The wavelength of the carrier phase is  $\lambda_i$. The float ambiguity at frequency $i$ is denoted by $N^s_{r,i}$. The noise of the pseudorange and carrier phase measurement is $e^s_{r,i}$ and $\epsilon_{r,i}^s$, with the variance of $\sigma^2_{P^s_{r,i}}$ and $\sigma^2_{L_{r,i}^s}$, respectively.

The GNSS observations from the base station help to significantly improve the positioning accuracy of the rover station. Satellite orbit and clock biases, ionospheric and tropospheric delays in (\ref{equ:gnss observation}) are eliminated by performing single-difference between the base station $b$ and rover station $r$ when the baseline is less than 10km.  The following is how we get the single-difference measurement model:

\begin{equation}
    \begin{aligned}
    \Delta P^s_{r,i}&= \Delta \bm{\rho}+c \Delta t_r + \Delta e_{r,i}^s,\Delta \sigma^2_{P^s_{r,i}} \\  
     \Delta L_{r,i}^s&=\Delta \bm{\rho}+c \Delta t_r +  \lambda_i \cdot \Delta N_{r,i}^s+ \Delta \epsilon_{r,i}^s, \Delta \sigma^2_{L_{r,i}^s}
    \end{aligned} 
\end{equation}
where $\Delta\left(\cdot\right)$ represents the single-difference operator.

We also introduce the IFB \cite{psychas2020real} for the RTK model with multi-system and multi-frequency observations. Subsequently, the state vector of single-difference RTK can be expressed as follows:
\begin{equation}
    \delta \mathbf{x}_{rtk} = \left[\delta \Delta \mathbf{p}^n_{r} \ \delta\Delta {t}^G_{r} \    \delta \Delta \mathbf{IFB}_{r,i} \  \delta \Delta \mathbf{N}_{r,i}^s\right]^{\top}\label{equ: difference gnss obs}
\end{equation} 
where $\delta\Delta \mathbf{p}^n_{r}$ indicates the error state of the baseline from a base station to a rover station. $\delta\Delta {t}^G_{r}$ is the error state of the single-difference GPS receiver clock, i.e., the datum receiver clock.  $ \delta \Delta \mathbf{IFB}_{r,i}$ is the error state of the single-difference IFB. And $\delta\Delta \mathbf{N}_{r,i}^s$ represents the error state of the single-difference ambiguities. 

It is worth noting that we estimate the single-difference float ambiguity in (\ref{equ: difference gnss obs}). After the measurement update at each epoch, we will fix the between-satellite between-station ambiguities to integers, if possible, then the fixed solutions will be obtained. In this study, the integer least-square estimation of the float ambiguities is searched using the least-square ambiguity decorrelation adjustment (LAMBDA) method \cite{teunissen1995lambda}. Although fixing ambiguities can enhance the positioning accuracy of RTK, if the float ambiguities are excessively erroneous, they tend to result in incorrect fixes, leading to fixed positioning solutions worse than the float solutions. Therefore, reliable initial float ambiguities are crucial for high-precision RTK positioning. 

\subsection{Tightly Coupled Filter Model for RTK/MEMS/Vision}

The MSCKF is employed to estimate the real-time state variables with single-difference observations of GNSS, raw observations of MEMS IMU, and the stereo camera. To realize the tightly coupled estimation, all the corresponding state variables must be held in one estimator, which is different from loosely coupled filter \cite{wang2023measurement}  or semi-tightly coupled filter \cite{xu2023unified}. The state variables in the tightly coupled filter are defined as:
\begin{equation}
    \delta \mathbf{x}= \left[\delta \mathbf{x}_{ins} \ \delta \mathbf{x}_{vis} \ \underbrace{ \delta \Delta t_{r}^G \   \delta \Delta \mathbf{IFB}_{r,i} \  \delta \Delta \mathbf{N}_{r,i}^s}_{\delta\mathbf{x}_{rtk}} \right]^{\top}
\end{equation} 

The  Kalman filter's time updates and measurement updates are used to estimate the optimal state variables. In the time updates, the state prediction and error state covariance are propagated. The INS states are propagated forward by INS mechanization. The RTK-related state variables and the camera poses in the sliding window are regarded as constant without process noises. Regarding the error state covariance, the error state variables are propagated using a continuous system model, which is provided by: 

\begin{equation}
\left[\begin{matrix}
    \delta \dot{\mathbf{x}}_{ins} \\ 
    \delta \dot{\mathbf{x}}_{vis} \\ \delta \dot{\mathbf{x}}_{rtk} 
    \end{matrix}\right] = 
    \left[\begin{matrix}
    \mathbf{F}_{ins} & \mathbf{0} & \mathbf{0} \\ 
    \mathbf{0} & \mathbf{0} & \mathbf{0}  \\
    \mathbf{0} & \mathbf{0} & \mathbf{0} 
    \end{matrix}\right] \left[\begin{matrix}
    \delta {\mathbf{x}}_{ins} \\ 
    \delta {\mathbf{x}}_{vis} \\ \delta {\mathbf{x}}_{rtk} 
    \end{matrix}\right] + \left[\begin{matrix}
    {\mathbf{n}}_{ins} \\ 
    {\mathbf{0}} \\ 
    {\mathbf{n}}_{rtk} 
    \end{matrix}\right] \label{equ: INS time update}
\end{equation} 
where the continuous-time state transition matrix of INS is represented by $\mathbf{F}_{ins}$. The process noise of INS is $\mathbf{n}_{ins}$, including the Gaussian noise of the accelerometer and gyroscope. And $\mathbf{n}_{rtk}$ is the process noise of RTK, containing the Gaussian noise of the receiver clock.

\begin{figure}[t]
  \centering
  \vspace{-1mm}
  \includegraphics[width=0.48\textwidth]{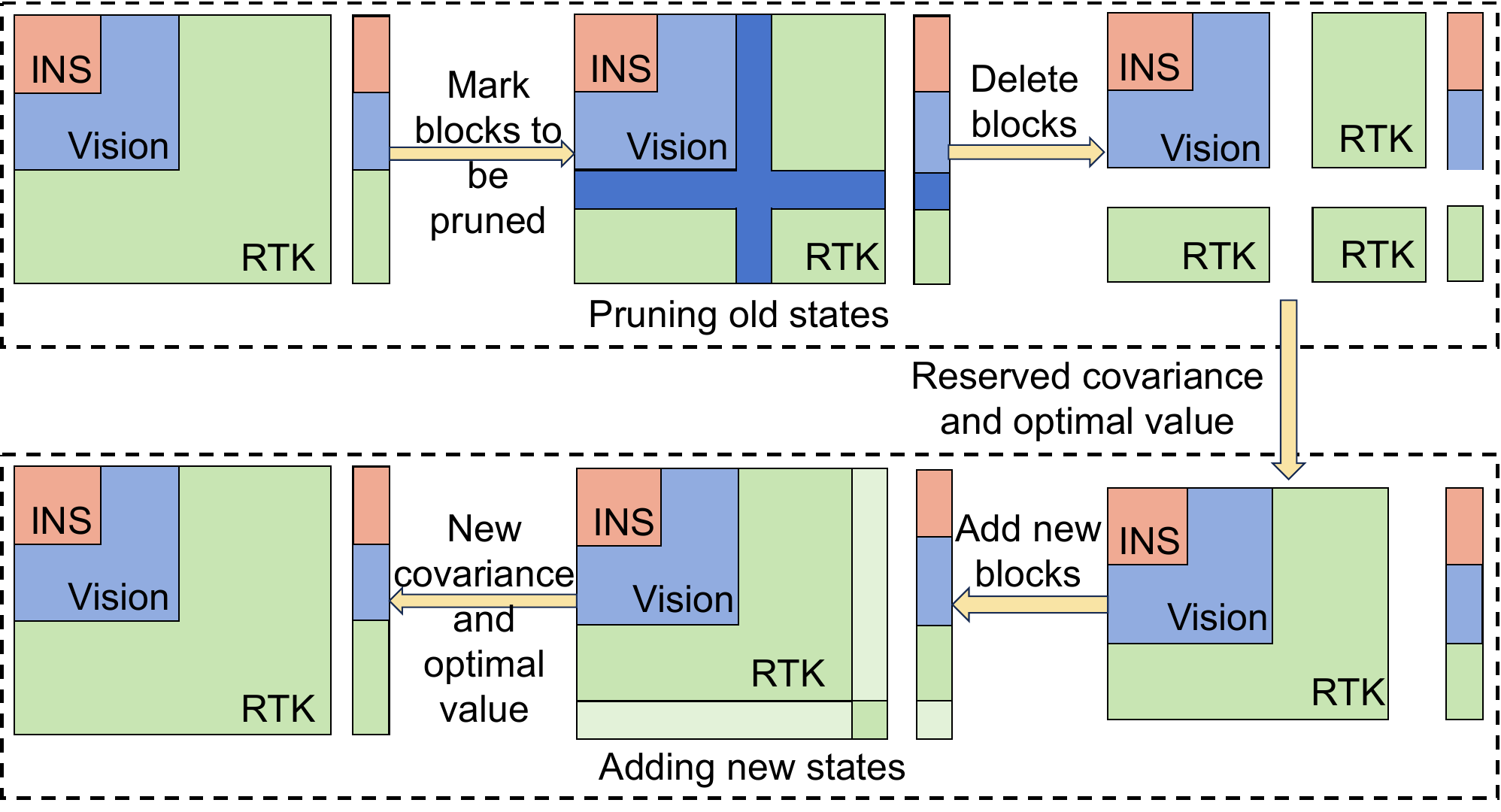}
  \vspace{-3mm}
  \caption{The augmentation of the optimal value and covariance when RTK observation is recorded. The optimal value and covariance from the newly added and that copied from the previous moment together constitute the optimal value and covariance of the current moment. }
  \label{fig:rtk_augement}
  \vspace{-4mm}
\end{figure}

The discrete form of the covariance propagation based on (\ref{equ: INS time update}) can be written as: 

\begin{equation}
    \mathbf{P}_{k} = \mathbf{\Phi}_{k, k-1} \mathbf{P}_{k-1}\mathbf{\Phi}_{k, k-1}^{\top} + \mathbf{Q}_{k-1}
\end{equation}
where $\bm{\Phi}_{k,k-1}$ is the discrete-time state transition matrix. $\mathbf{Q}_{k-1}$ is the discrete-time noise covariance matrix. The covariance matrix and state variables are augmented each time the new visual or RTK observation is captured. RTK augmentation is simple. The optimal value and covariance are carried over from the previous epoch if the state variables from that epoch are present in the current epoch, as seen in Fig. \ref{fig:rtk_augement}. For the newly added state variables in the current epoch, the optimal error state value is set to $\mathbf{0}$, and the variance is set as Gaussian noise. The INS mechanization would be used to initialize the camera pose for the new add image, and the augmented covariance can be written as follows:

\begin{equation}
\mathbf{P}'_{k} = \left(\begin{matrix}
    \mathbf{I}_{15+\gamma + 6k} \\ \mathbf{H}
\end{matrix}\right)  \mathbf{P}_{k} \left(\begin{matrix}
    \mathbf{I}_{15+\gamma + 6k} \\ \mathbf{H}
\end{matrix}\right)^{\top}
\end{equation}
with $\mathbf{H} = \left[\begin{matrix}
\mathbf{R}{^b_c}^{\top} & \mathbf{0} & \mathbf{0} & \mathbf{0}& \mathbf{0} & \mathbf{0}_{3 \times (\gamma + 6 k)} \\ -\mathbf{R}_b^n (\mathbf{p}^b_c)^{\wedge} & \mathbf{0} & \mathbf{I} & \mathbf{0} & \mathbf{0}  & \mathbf{0}_{3 \times (\gamma + 6 k)}
\end{matrix}\right]$. $\gamma$ and $k$ are the number of the RTK-related parameters and camera poses.  $\mathbf{R}_c^b$ and $\mathbf{p}_c^b$ are the offline-calibrated extrinsic parameters that connect the IMU and camera \cite{rehder2016extending}. 

The tightly coupled measurement update can 
be represented as follows when the GNSS or visual measurements are available: 

\begin{equation}
\begin{aligned}
   \left[ \begin{matrix}
        \mathbf{r}_{vis} \\ \mathbf{P}^s_{r,i} \\ \mathbf{L}^s_{r,i}
    \end{matrix} \right] &=  \left[ \begin{matrix}
        \mathbf{z}_{vis} - \mathbf{\hat{z}}_{vis} \\ \Delta \mathbf{P}^s_{r,i} -  \Delta \mathbf{\hat{P}}^s_{ins,i} \\ \Delta \mathbf{L}^s_{r,i} -  \Delta \mathbf{\hat{L}}^s_{ins,i}
    \end{matrix} \right]  \\
    &=  \left[ \begin{matrix}
        \mathbf{H}_{vis} \\ \mathbf{H}_{\Delta \mathbf{P}^s_{r,i}} \\\mathbf{H}_{\Delta \mathbf{L}^s_{r,i}}
    \end{matrix} \right]  \left[ \begin{matrix}
        \mathbf{x}_{vis} \\ \delta \mathbf{x}_{vis} \\ \delta \Delta t^G_{r} \\ \delta \Delta \mathbf{IFB}_{r,i}\\
        \delta \Delta \mathbf{N}_{r,i}^s
    \end{matrix} \right] +  \left[ \begin{matrix}
        \mathbf{n}_{vis} \\ \Delta \mathbf{e}^s_{r,i} \\ \Delta \bm{\epsilon}^s_{r,i}
    \end{matrix} \right]
    \end{aligned}
\end{equation}
where  $\Delta \mathbf{\hat{L}}^s_{ins,i}$ and $\Delta \mathbf{\hat{P}}^s_{ins,i}$ represent the INS-predict single-difference GNSS carrier phase measurement and the pseudorange measurement, respectively. $\mathbf{H}_{vis}$, $\mathbf{H}_{\Delta \mathbf{P}^s_{r,i}}$ and $\mathbf{H}_{\Delta \mathbf{L}^s_{r,i}}$ represent the Jacobian matrix of vision representation error, pseudorange error and carrier phase error, respectively. Since the INS central position $\mathbf{p}_{ins}^n$ does not overlap with the GNSS receiver antenna reference point $\mathbf{p}_{rtk}^n$, we also consider the lever-arm correction $\mathbf{l}^b$ in our study.

\begin{figure}[t]
  \centering
  \vspace{-1mm}
  \includegraphics[width=0.48\textwidth]{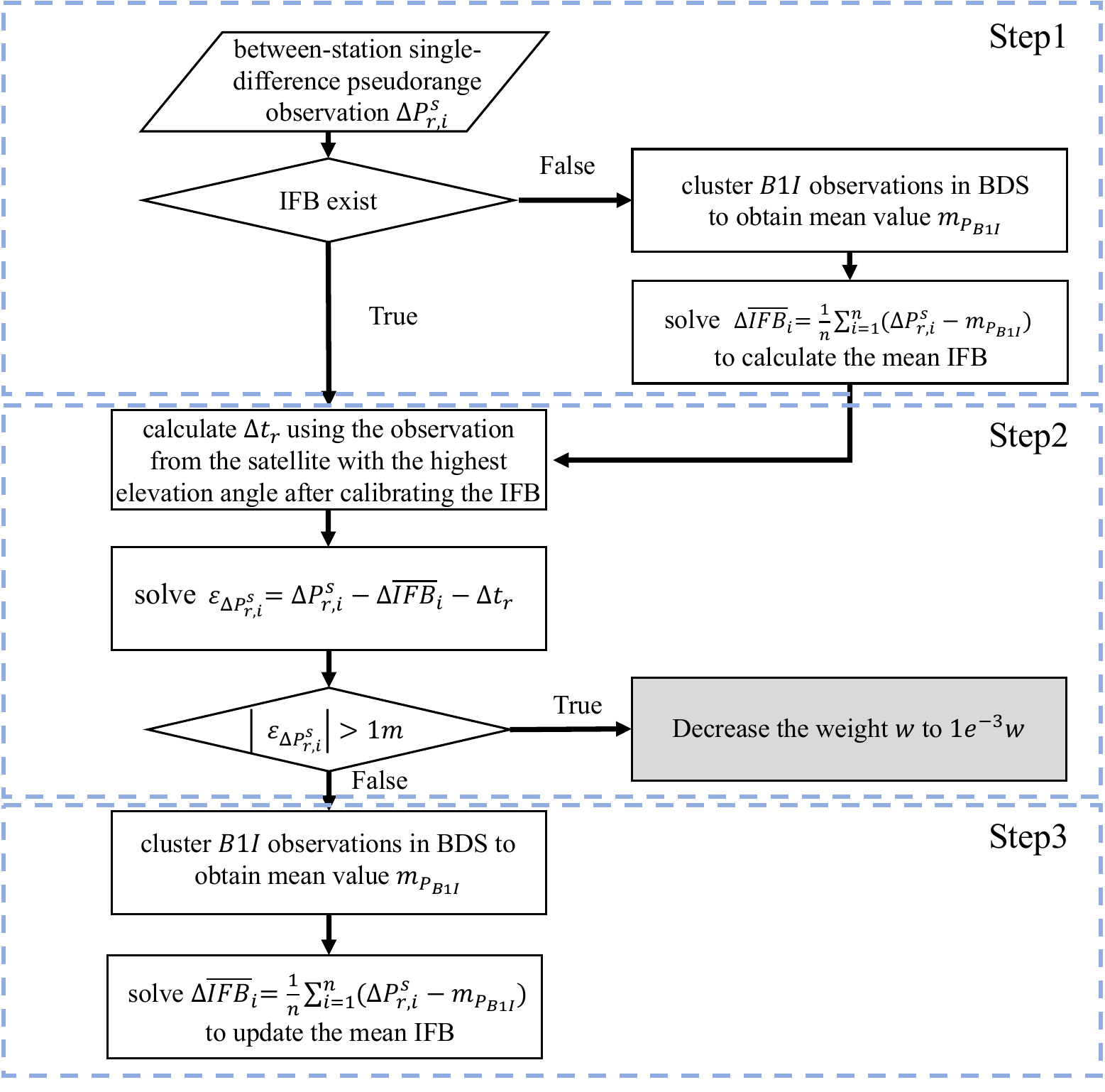}
  \vspace{-3mm}
  \caption{Implementation of pseudorange processing in I-EDM algorithm. The algorithm utilizes the innovation to detect the multipath in the observations, thereby avoiding the estimation of the parameters with least square method.}
  \label{fig:pseudorange_process}
  \vspace{-4mm}
\end{figure}

\subsection{Innovation-based EDM Method}
\begin{figure}[t]
  \centering
  \vspace{-1mm}
  \includegraphics[width=0.48\textwidth]{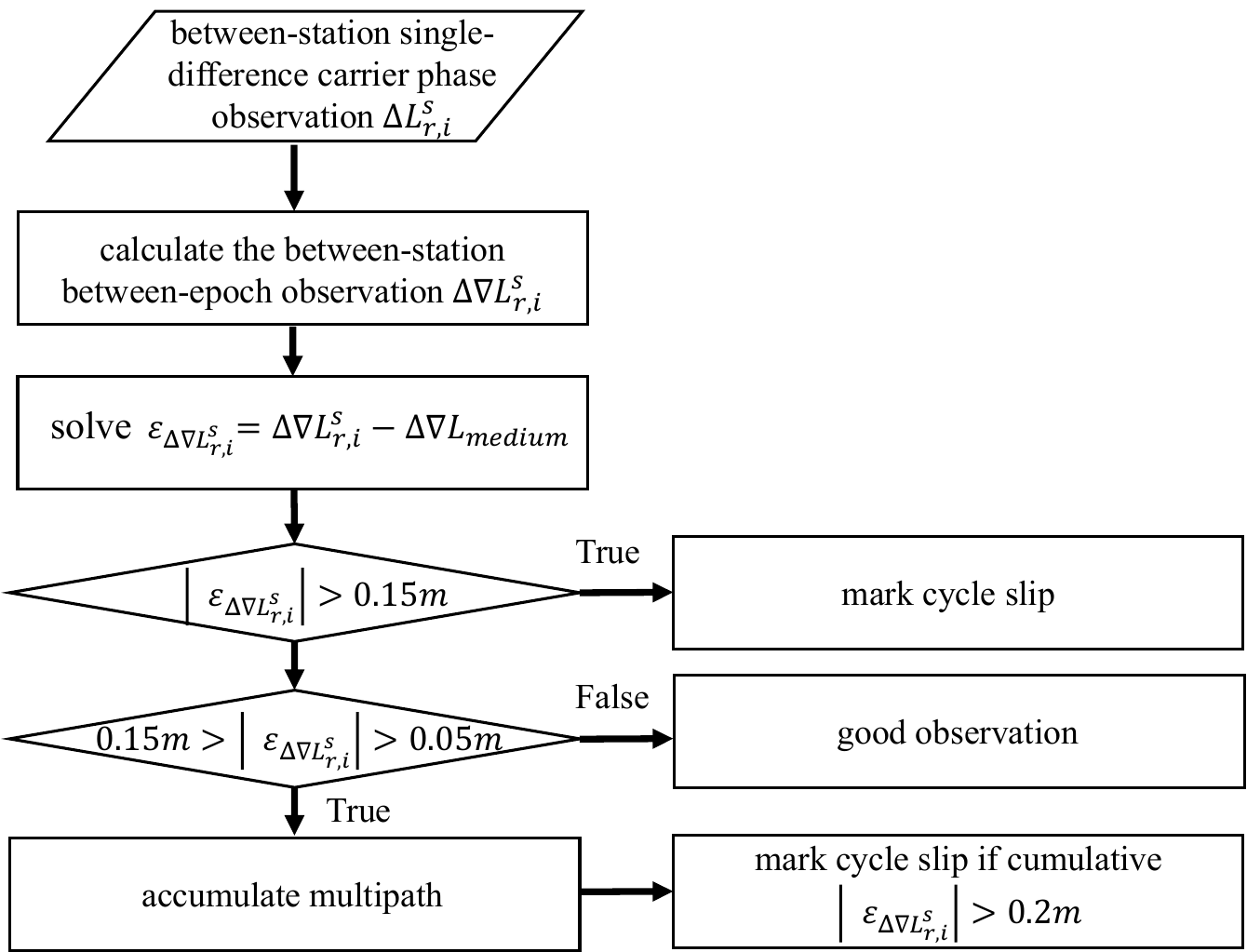}
  \vspace{-3mm}
  \caption{Implementation of carrier phase processing in I-EDM algorithm. The algorithm detects cycle slips and models multipath with between-station between-epoch carrier phase observations. }
  \label{fig:carrier_process}
  \vspace{-4mm}
\end{figure}

Based on the hybrid model of SNR and satellite elevation angle \cite{herrera2016gogps}, we propose an innovation-based EDM method for GNSS pseudorange and carrier phase observations, which avoids residual overfitting with the least square method. As illustrated in Fig.\ref{fig:pseudorange_process} and Fig.\ref{fig:carrier_process}, the pseudorange and carrier phase data are processed independently by the innovation-based EDM approach. For pseudorange processing, we assume all the pseudorange observations share the same receiver clock $\Delta {t}_r$, and the observations from different frequencies and systems are modeled with IFB. There are three steps in the pseudorange processing algorithm:

\begin{itemize}
  \item [1)] 
 IFB initialization: at the initial epoch, IFB needs to be computed first. We choose the code observations on $B1I$ frequency of the BDS system as the reference observations, which are the most among all the observations. The DBSCAN cluster analysis method \cite{ester1996density} is utilized to obtain the stable mean value $m_{P_{B1I}}$. Then we clustered the observations from different frequencies after calibrating with $m_{P_{B1I}}$. The mean IFBs of different frequencies are computed from the most groups in the clustering results: 
    
\begin{equation}
    \Delta \overline{IFB}_i = \frac{1}{n} \sum^n_{i=1}\left(\Delta P^s_{r,i} - m_{P_{B1I}}\right) \label{equ: calculate ifb}
\end{equation}

  \item [2)]
   Outlier culling: we assume the observations of satellites with the highest elevation angle are almost not affected by the multipath effect. Then the receiver clock $\Delta {t}_r$ is computed with these observations after calibrating the IFB.  After obtaining  $\Delta \overline{IFB}_i$ and $\Delta {t}_r$, innovations of different observations, i.e. the prefit-residuals, can be computed by:  
\begin{equation}
    {\mathcal{E}}_{\Delta P^s_{r,i} } = \Delta P^s_{r,i} - \Delta\overline{IFB}_i - \Delta t_r
\end{equation}
Finally, the weights of innovations greater than 1m are decreased from $w$ to $1e^{-3}w$. The reason for choosing pseudorange values greater than 1m as outliers is that the standard deviation of pseudorange is set to 0.3m in our research. Therefore, the innovation which is greater than 3 times standard deviation needs to be culled.  
  \item [3)]
  IFB update: we compute the $m_{P_{B1I}}$ and update the mean IFBs of different frequencies as (\ref{equ: calculate ifb}) in each epoch to obtain the robust estimation of IFB. 
\end{itemize}

For carrier phase observations, the algorithm is simplified due to the adoption of between-station between-epoch double difference carrier phase observations $\Delta \nabla L^s_{r,i}$.
In order to determine the innovation of the observations, we further assume that all the carrier phase observations share the same receiver clock, with medium value serving as the datum. Next, using the following criteria \cite{xu2023unified}, the observations that belong to the good, multipath, and cycle slips are identified.

\begin{equation} \label{equ: threshold of EDM}
    f\left(\mathcal{E}_{\Delta \nabla L^s_{r,i}} \right)=\left\{
\begin{aligned}
 good,&   & |\mathcal{E}_{\Delta \nabla L^s_{r,i}}| < \delta_{th} \\
multipath,&  &  \delta_{th} < |\mathcal{E}_{\Delta \nabla L^s_{r,i}}| <  3 \cdot \delta_{th} \\
cycle slip,& & |\mathcal{E}_{\Delta  \nabla L^s_{r,i}}| >  3 \cdot \delta_{th}
\end{aligned}
\right.
\end{equation}
$\delta_{th}$ is set to $0.05m$ in our experiments. It is worth noting that the biased ambiguities will still have an impact on the system's positioning performance if the multipath is absorbed by the ambiguities during the Kalman filter. The multipath will be gathered and designated as the cycle slip if the aggregated value is above $0.2m$ to mitigate the multipath effect on ambiguity. 

\section{Experiments}
\begin{figure}[t]
  \centering
  \vspace{-1mm}
  \includegraphics[width=0.48\textwidth]{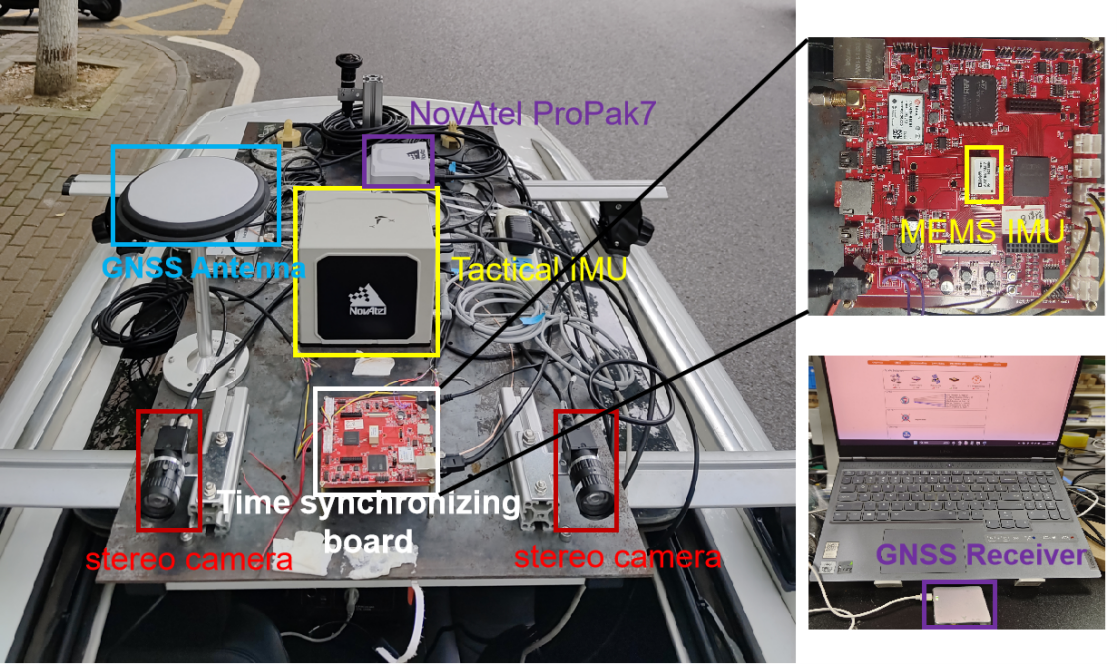}
  \vspace{-3mm}
  \caption{ The hardware equipment used for data collection. The GNSS/INS/Vision experimental data is collected with a stereo camera, a MEMS IMU, a tactical IMU, and a GNSS antenna. }
  \label{fig:equipment}
  \vspace{-4mm}
\end{figure}

Road vehicular experiments in urban regions are carried out in Wuhan, China, to assess the performance of the tightly coupled RTK/IMU/Vision system with the proposed I-EDM method. As seen in Fig. \ref{fig:equipment}, the experimental vehicle was outfitted with two FLIR BFS-U3-31S4C-C cameras,  a low-cost ADIS-16470 MEMS IMU, a tactical-grade NovAtel SPAN-ISA-100C IMU, a time synchronization board and a Septentrio mosaic-X5 mini GNSS receiver with a NovAtel GNSS-850 antenna.  The vehicle-borne mobile system aboard the vehicle was used to collect the experimental data. Tab.\ref{tab: equipment parameter} contains the specification information of the consumer-grade and tactical-grade IMUs.

\begin{table*}[t!]
	\captionsetup {font={small,stretch=1.5}}
	\caption{Technical specifications of the IMU sensors.} 
	\label{tab: equipment parameter}
	\begin{center}
	\setlength\tabcolsep{15pt}
	\setlength{\belowcaptionskip}{1pt}
    \renewcommand\arraystretch{1.0}
	\small
 	\vspace{-2.0em}
	\resizebox{\linewidth}{!}{
		\begin{tabular}{ccccccc}
			\toprule
			  \multirow{2}{*}{IMU Equipment} & \multirow{2}{*}{Grade} & \multirow{2}{*}{Sample rates (Hz)} & Angular & Velocity & Acc & Gyro\\
             &  &  & $\left(^\circ/\sqrt{h}\right)$ & $\left(m/s/\sqrt{h}\right)$ & $\left(mGal\right)$ & $\left(^\circ/h\right)$\\
			\midrule
	ADIS-16470  & MEMS & 100 & 0.34 & 0.18 & 1300 & 8 \\
        SPAN-ISA-100C  & Tactical & 200 & 0.005 & 0.018 & 100 & 0.05  \\
             \specialrule{0.08em}{1pt}{1pt}
		\end{tabular}
		\vspace{-2em}
		}
	\end{center}
\end{table*}

Furthermore, a single  Septentrio PolaRx5 GNSS receiver was installed a mere 5$km$ away from the rover station, which functions as the base station with exact coordinates. The single-difference GNSS measurements between the base and rover stations are produced. We used the commercial Inertial Explorer (IE) 8.9 software \cite{Label-IE} to acquire the accurate smoothed solutions of the tightly coupled multi-GNSS post-processing kinematic (PPK) and tactical-grade IMU integration, with which the experimental results are verified. When
compared to MEMS-IMU, the tactical-grade IMU can sustain a particular level of pose output
accuracy for a comparatively extended period of time without the need for external corrective data, ensuring a trustworthy comparison in the assessment that follows.

\begin{table*}[t!]
	\captionsetup {font={small,stretch=5.5}}
	\caption{Specific parameters of our tightly coupled GNSS/INS/Vision system} 
	\label{tab: system parameter}
	\begin{center}
	\setlength\tabcolsep{10pt}
	\setlength{\belowcaptionskip}{1pt}
    \renewcommand\arraystretch{1.0}
	\small
 	\vspace{-2.0em}
	\resizebox{\linewidth}{!}{
		\begin{tabular}{c|c}
			\toprule
			  Items & Processing or estimation parameter \\
			\midrule
	Vision sliding window size  & 20\\
        Vision feature observation noise & 0.05m\\
        Image resolutions  & 1024 $\times$ 768 pixels\\
        Vision sampling rate  & 10Hz\\
        Maximum tracking features  & 150\\
        IMU forward propagation  & Fourth-order Runge‒Kutta integrator\\
        GNSS systems  & GPS, BDS, Galileo, QZSS\\
        GNSS signal selection  & GPS:L1, L2, L5; Galileo: E1, E5a, E5b; BDS: B1, B2, B3; QZSS:L1, L2, L5\\
        Satellite elevation cutoff  & 10$^\circ$\\
        Satellite orbit and clock  & Real time products from the Centre National d’Etudes Spatiales (CNES) \cite{laurichesse2011cnes}\\
        Satellite antenna phase center  & Corrected with IGS14.ATX\\
        Phase windup  & Corrected \cite{wu1992effects}\\
        Pseudorange observation noise   & 0.3m\\
        Carrier phase observation noise   & 0.03m\\
        IFB   & Estimated\\
        Ambiguity   & Fixed\\
       
             \specialrule{0.08em}{1pt}{1pt}
		\end{tabular}
		\vspace{-2em}
		}
	\end{center}
\end{table*}
\begin{figure}[b]

  \centering
  \vspace{-1mm}
  \includegraphics[width=0.48\textwidth]{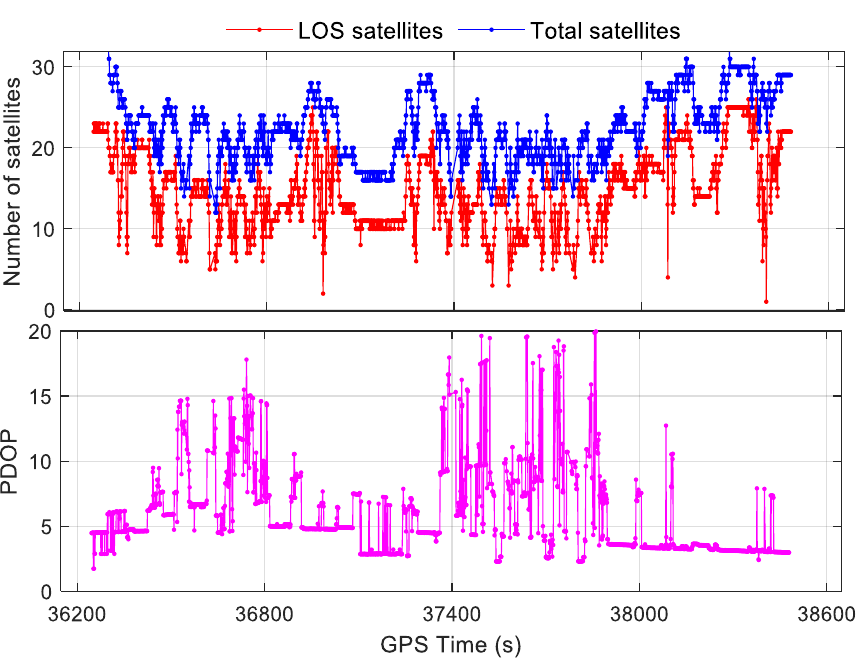}
  \vspace{-3mm}
  \caption{Number of available satellites (top) and PDOP (bottom) in the urban scenarios.}
  \label{fig:pdop_num_sat}
  \vspace{-4mm}
\end{figure}

\begin{figure}[t]
  \centering
  \vspace{-1mm}
  \includegraphics[width=0.48\textwidth]{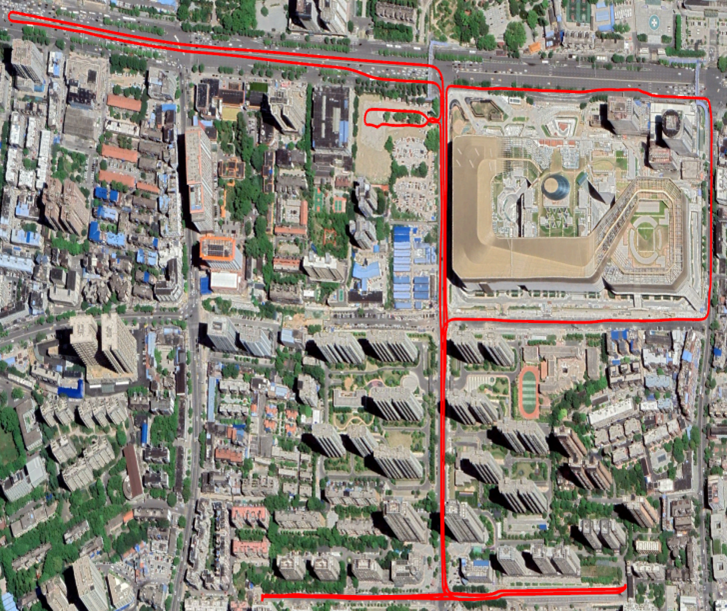}
  \vspace{-3mm}
  \caption{ Google Earth's 
  Top view of the vehicle trajectory from Google Earth. Our experimental scene includes typical urban features such as forests and tall buildings.}
  \label{fig:top view trajectory}
  \vspace{-4mm}
\end{figure}

\begin{figure}[t]
  \centering
  \vspace{-1mm}
  \includegraphics[width=0.48\textwidth]{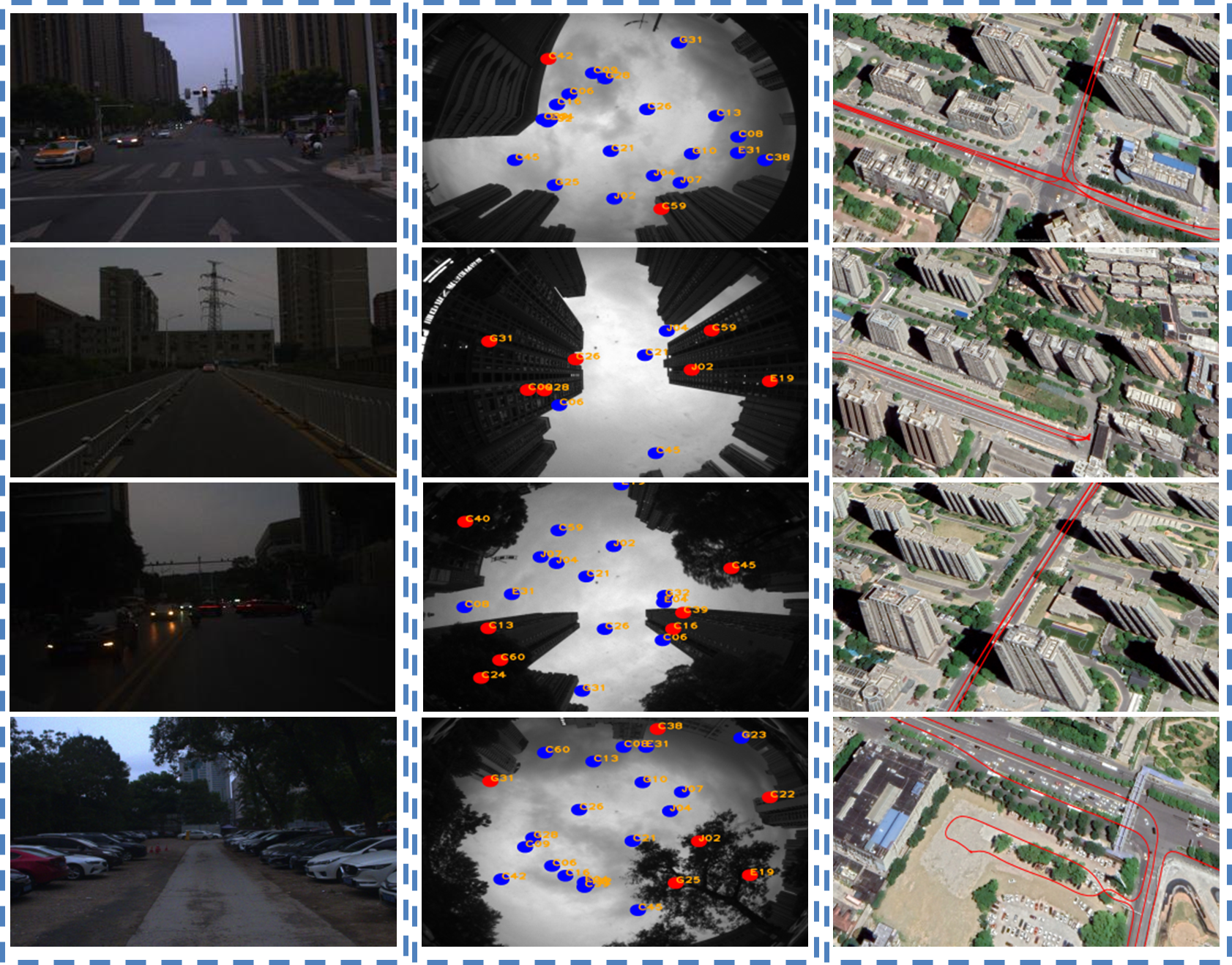}
  \vspace{-3mm}
  \caption{Typical experimental scenes (left), skyplots (middle) and vehicle trajectories overlaid on Google Maps (right) in the urban area. The blue and red dots in the skyplots represent the LOS and NLOS satellites, respectively.}
  \label{fig:typical_experimental_scenes}
  \vspace{-4mm}
\end{figure}

The vehicular dataset, including both open-sky and urban settings, is collected between 18:04:06 and 18:41:19 on September 3, 2023, is used for a complete evaluation. Fig. \ref{fig:pdop_num_sat} displays the number of available satellites and the position dilution of precision (PDOP). The average number of LOS, NLOS, and total satellites are 14.57, 7.77, and 22.34. The average number of available satellites for GPS, BDS, Galileo, and QZSS satellite system is 4.25, 11.86, 3.64, and 2.59, respectively. The PDOP has an average value of 5.84. There is a sharp decline in the number of LOS satellites while the car is traveling between towering structures. The experimental scenarios are severely affected by GNSS NLOS, multipath effect, and cycle slip issues, negatively impacting the RTK/INS/Vision integration system's positioning performance. The trajectory's top view is shown in Fig. \ref{fig:top view trajectory}. Additionally, the vehicle trajectory's typical experimental situation are depicted in Fig. \ref{fig:typical_experimental_scenes}. The trajectory's overall length is roughly 8587$m$. Tab. \ref{tab: system parameter} displays the specific parameters for the GNSS, INS, and Vision modules. The processing frequency for GNSS, INS, and Vision is 1$HZ$, 100$HZ$, and 10$HZ$, respectively. The experiments are carried out on a PC with an Intel Core i7-9750 @ 2.6 GHz and 16 GB of RAM.
\subsection{Performance of Different Weighting Strategies}
\begin{figure}[b]
  \centering
  \vspace{-1mm}
  \includegraphics[width=0.5\textwidth]{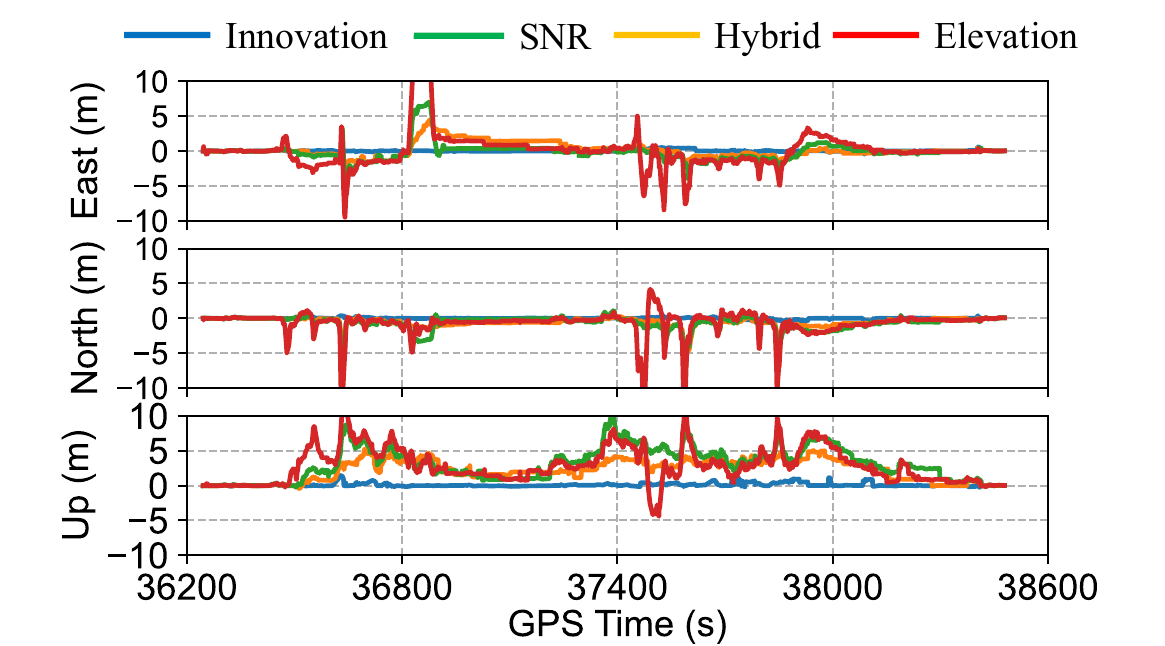}
 
  \caption{Positioning accuracy comparison of SNR, satellite elevation angle method, SNR and satellite elevation angle hybrid method and innovation-based preprocessing method in urban canyon scenarios}
  \label{fig:diff_weight_traj}
  
\end{figure}

In urban canyon scenarios, the GNSS observations are susceptible to multipath and outliers, consequently decreasing the positioning performance of the GNSS/INS/Vision integration system. Therefore, choosing appropriate weights is essential for GNSS observations. We compared the positioning performance of the current mainstream weighting methods of GNSS pseudorange observations, including signal-to-noise ratio (SNR) model \cite{hartinger1999variances}, $1/{\sin\theta}^2$ satellite elevation angle model \cite{gao2011improved}, SNR and satellite elevation angle hybrid model \cite{herrera2016gogps} and the proposed I-EDM method in RTK/INS/Vision integration system. The trajectory errors in east, north, up for SNR, satellite elevation angle, hybrid method, and innovation-based method are displayed in Fig. \ref{fig:diff_weight_traj}. The root mean square error (RMSE) of the trajectory error is computed and displayed in Tab.\ref{tab: accuracy_diff_weight_traj}. We can observe that the proposed I-EDM method achieves the highest positioning accuracy of 0.31$m$. This is because our method effectively models the multipath effects in the GNSS signals by computing the innovation of the pesudorange observations and carrier phase observations. Therefore the accuracy of the positioning is significantly improved by eliminating the impact of the multipath effects. On the contrary, the other three methods do not account for the significant multipath effect caused by the obstructions in urban situations, resulting in the positioning accuracy only reaching the meter level. Then, by comparing the hybrid method with the SNR and satellite elevation angle method, we can see that the positioning performance of the hybrid method is better than that of the SNR and satellite elevation angle method, achieving on average a 28.1\% and 38.4\% improvement in 3D direction. And the performance of the SNR method is better than that of the satellite elevation angle method. The SNR method allows for assigning reasonable weights to the observations of different frequencies and types from the same satellite. However, the satellite elevation angle method only assigns the weights according to the satellite type. Consequently, the SNR method provides more accurate modelling of observations at the signal level, resulting in higher precision in positioning compared to the satellite elevation angle method. Furthermore, the hybrid method combines the advantages of both methods, achieving better positioning accuracy. 

\subsection{Comparison of R-EDM and I-EDM Method}
We conduct experiments to evaluate the performance of R-EDM and I-EDM in terms of cycle slip detection, multipath estimation, positioning accuracy, and running time consumption. The aim is to provide a detailed analysis of the differences between the residual-based EDM method and the innovation-based  EDM  method in urban environments.

\begin{table}[t!]
	\captionsetup {font={small,stretch=5.5}}
	\caption{RMSEs of the position errors for SNR, satellite elevation angle method, SNR and satellite elevation angle hybrid method, and innovation-based preprocessing method} 
	\label{tab: accuracy_diff_weight_traj}
	\begin{center}
	\setlength\tabcolsep{10pt}
	\setlength{\belowcaptionskip}{1pt}
    \renewcommand\arraystretch{1.0}
	\small
 	\vspace{-2.0em}
	\resizebox{\linewidth}{!}{
		\begin{tabular}{c|cccc}
			\toprule
			 & East(m) & North (m)& Up(m) & 3D\\
			\midrule
	SNR  & 1.31 & 1.19 & 3.85 & 4.24 \\
        Elevation  & 2.60 & 1.85 & 3.79 & 4.95  \\
        Hybrid  & 1.18 & 0.93 & 2.65 & 3.05  \\
        Innovation  & 0.12 & 0.11 & 0.26 & 0.31  \\
             \specialrule{0.08em}{1pt}{1pt}
		\end{tabular}
		\vspace{-2em}
		}
	\end{center}
\end{table}
\subsubsection{Cycle Slip Detection Results}
\begin{figure}[b]
  \centering
  \vspace{-1mm}
  \includegraphics[width=0.5\textwidth]{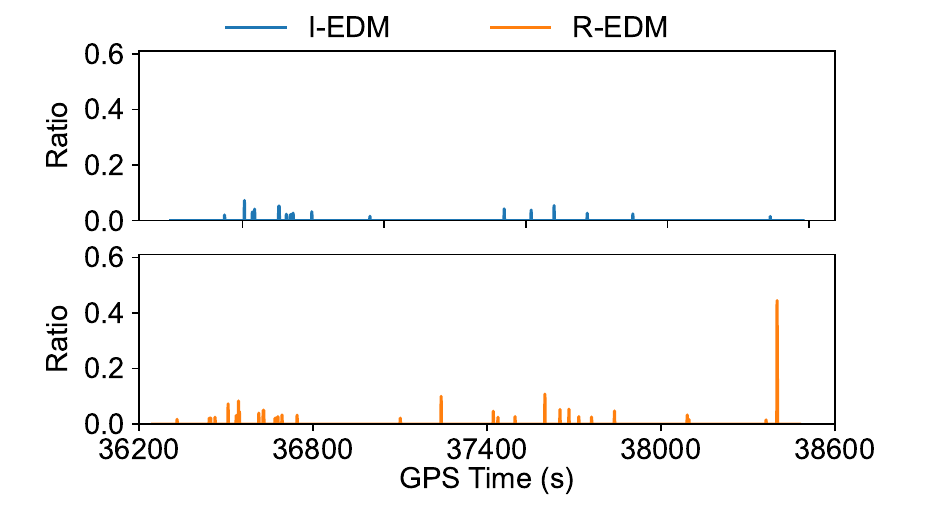}
 
  \caption{The false detection rate of cycle slip using I-EDM and R-EDM methods in the urban situations}
  \label{fig:cycle_slip}
  
\end{figure}

We compared the performance of the cycle slip detection of R-EDM and I-EDM methods. The reference trajectory generated from the tactical-grade IMU is employed to obtain the real cycle slips and multipath \cite{xu2023unified}. For fairness, the same criteria described in equation (\ref{equ: threshold of EDM}) are utilized to access the cycle slips and multipath of R-EDM, I-EDM, and real values. The false detection rate of cycle slip is displayed in Fig. \ref{fig:cycle_slip}. The false detection rate means that the cycle slips exist within the real data, yet the method fails to detect them. This omission leads to incorrect ambiguity calculation, which is harmful to navigation performance, especially in urban experiments. The periods when cycle slip detection errors occur essentially coincide with intervals of high PDOP values, as shown in Fig.\ref{fig:pdop_num_sat}. This correlation demonstrates that multipath effects and signal loss in GNSS observations often result in cycle slips. More specifically, R-EDM and I-EDM have average error rates of 0.07\% and 0.03\%, respectively. This demonstrates that in the residual-based method, the cycle slips are partially absorbed into the estimated parameters such as IFB and clock biases when solving least squares, making it challenging to detect the minor cycle slips within the GNSS observations. Conversely, the innovation-based method only leverages the raw GNSS observations to detect the cycle slips, which are more sensitive to minor cycle slips.

\begin{figure}[t]
  \centering
  \vspace{-1mm}
  \includegraphics[width=0.48\textwidth]{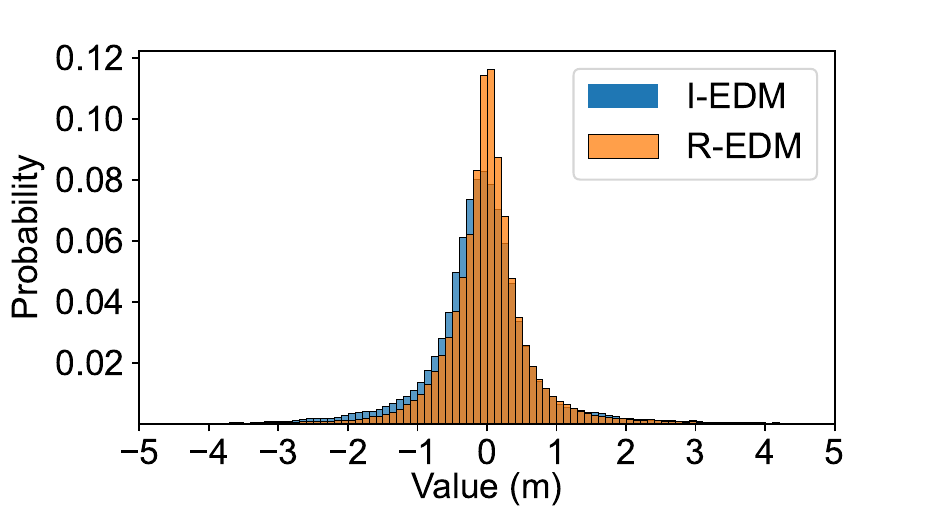}
 
  \caption{Probability distribution histogram of multipath error obtained by I-EDM and R-EDM method. The multipath exceeding $\pm1$m detected by I-EDM is greater than R-EDM.}
  \label{fig:pseudo_multipath}
  
\end{figure}

\begin{figure}[b]
  \centering
  \vspace{-1mm}
  \includegraphics[width=0.48\textwidth]{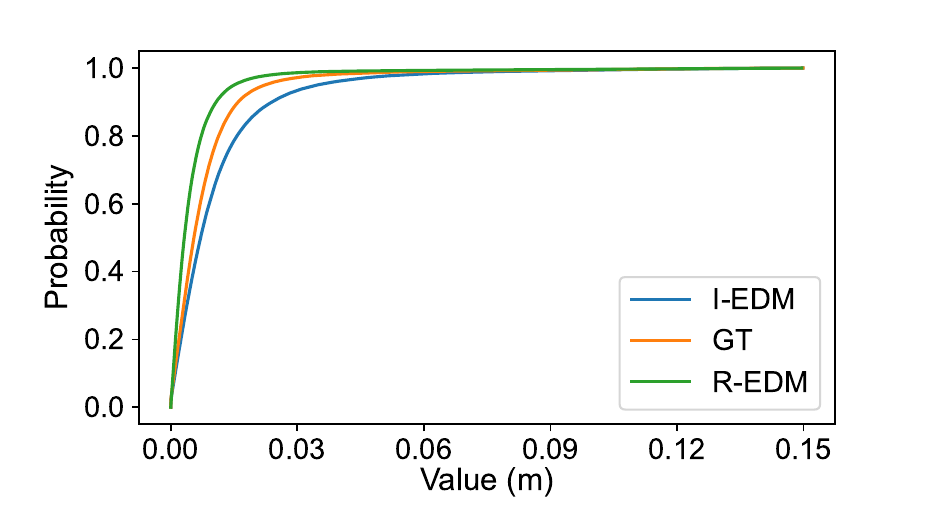}
 
  \caption{Cumulative distribution plots of multipath obtained by I-EDM, R-EDM, and groundtruth. The threshold of multipath estimation is shown in the X-axis, while the Y-axis indicates the fraction of the estimated multipath values below the corresponding threshold.}
  \label{fig:carrier_multipath}
  
\end{figure}

\begin{figure}[]
  \centering
  \vspace{-1mm}
  \includegraphics[width=0.5\textwidth]{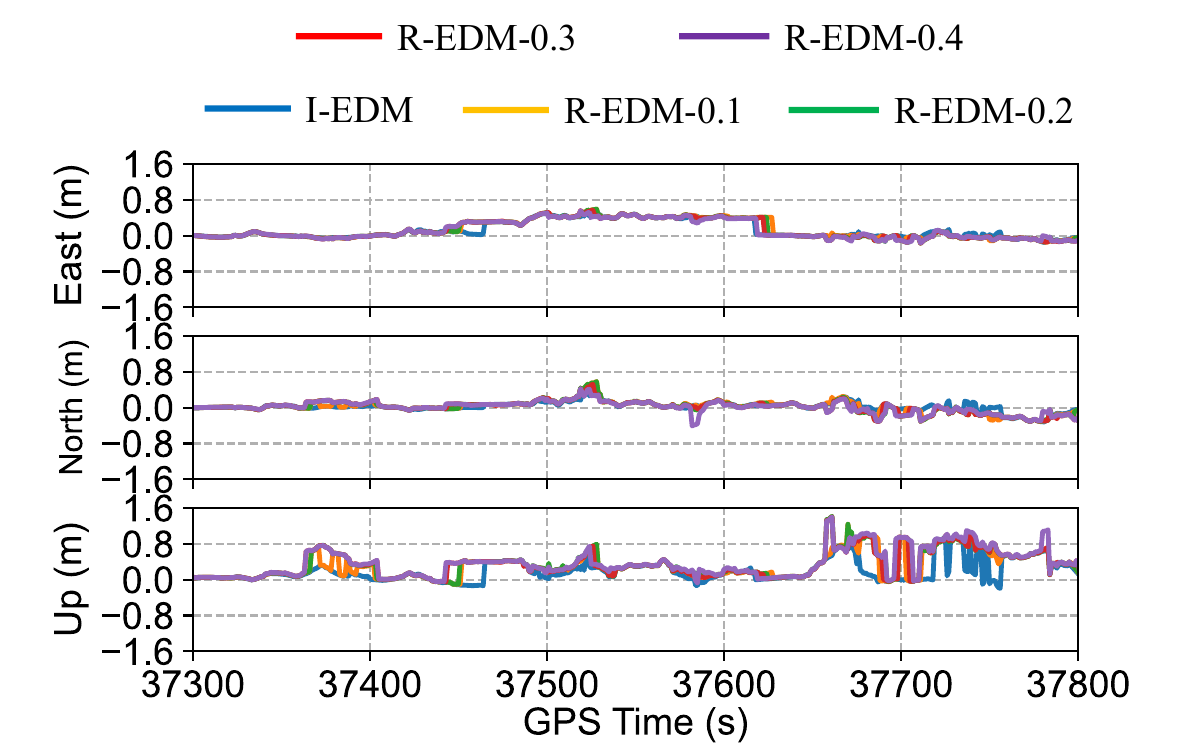}
  \caption{Trajectory error for the R-EDM method with different uncertainty of prior position constraints and I-EDM method for epochs from 37300 to 37800.}
  \label{fig:R-EDM-diff-weight}
\end{figure}
\subsubsection{Multipath Estimation Results}
We also compared the performance of the I-EDM and R-EDM methods for the multipath estimation of pseudorange and carrier phase observations. In Fig.\ref{fig:pseudo_multipath}, we plot the probability distribution histogram of the pseudorange multipath obtained by I-EDM and R-EDM methods. The multipath exceeding $\pm1$m  detected by I-EDM is greater than R-EDM method, which is  15.70\% and 11.87\%, respectively. This indicates that the innovation-based method can mitigate the impact of the multipath effect in pseudorange observations on the positioning system by
modeling the multipath effect effectively. The cumulative distribution of the multipath in carrier phase observations acquired by I-EDM, R-EDM, and the actual value is displayed in Fig.\ref{fig:carrier_multipath}. It can be observed that the multipath estimated by I-EDM is smaller than R-EDM. However, due to the between-station between-epoch double difference operator on carrier phase observations, the RMSE of the multipath estimated by I-EDM and R-EDM is similar, which is 0.011m and 0.009 m, respectively.

\begin{table}[t!]
	\captionsetup {font={small,stretch=5.5}}
	\caption{RMSEs of the position error for the R-EDM method with different uncertainty of prior position constraints and I-EDM method} 
	\label{tab: accuracy_diff_prior_traj}
	\begin{center}
	\setlength\tabcolsep{10pt}
	\setlength{\belowcaptionskip}{1pt}
    \renewcommand\arraystretch{1.0}
	\small
 	\vspace{-2.0em}
	\resizebox{\linewidth}{!}{
		\begin{tabular}{c|cccc}
			\toprule
			 & East(m) & North (m)& Up(m) & 3D\\
			\midrule
	R-EDM-0.1  & 0.25 & 0.14 & 0.42 & 0.51 \\
        R-EDM-0.2  & 0.24 & 0.14 & 0.46 & 0.54  \\
        R-EDM-0.3  & 0.24 & 0.13 & 0.45 & 0.53  \\
        R-EDM-0.4  & 0.23 & 0.13 & 0.48 & 0.55  \\
	I-EDM      & 0.23 & 0.11 & 0.31 & 0.40 \\
             \specialrule{0.08em}{1pt}{1pt}
		\end{tabular}
		\vspace{-2em}
		}
	\end{center}
\end{table}

\subsubsection{Positioning Results}
The positioning accuracy of different preprocessing methods is estimated in the tightly coupled RTK/MEMS/Vision mode. Due to the R-EDM method relying on the prior position constraints, we set the standard deviations of 0.1, 0.2, 0.3, and 0.4$m$ to the predicted position of VIO, and analyzed the impact on the R-EDM method. Tab. \ref{tab: accuracy_diff_prior_traj} displays the RMSE of the approximated trajectory errors. The position errors calculated with different uncertainty of the prior position show a maximum difference of 6$cm$ in the vertical direction. This indicates that the performance of the R-EDM method is influenced by the uncertainty of prior position constraints. When GNSS loses tracking frequently, setting the uncertainty of the prior position estimated by IMU and vision observations becomes challenging. The I-EDM method, on the other hand, avoids the problem of setting uncertainty of prior position constraints and the deficiency estimation of the outliers. The error trajectory of the I-EDM method decreases by 21.57\% compared with the R-EDM-0.1 method. Fig.\ref{fig:R-EDM-diff-weight} is the error trajectory for epochs from 37300-37800 with severe multipath effect, which presents the trajectory errors of different methods more clearly. The benefits of the innovation-based method are further validated.

\subsubsection{Running Time Results}
We counted the time consumption for the pseudorange process and carrier phase process module in the R-EDM and I-EDM algorithms. The computation cost (in milliseconds) of different modules is shown in Table \ref{tab: time consumption}.  In the R-EDM method, employing the clustering analysis avoids searching for the outliers in the observations with the iterative method, thereby decreasing the algorithm’s time consumption. However, the algorithm still introduces the least square method to compute the residuals of pseudorange and carrier phase observations, which consumes most of the time. In the I-EDM method, we only computed the innovation of the pseudorange and carrier phase observations. Compared with the R-EDM method, the pseudorange processing, carrier phase processing, and total processing time decreased by 46.5\%, 51.5\%, and 49.9\%, respectively.

\begin{table}[t!]
	\captionsetup {font={small,stretch=0.5}}
	\caption{mean execution time (unit: millisecond) of R-EDM and I-EDM methods.} 
	\label{tab: time consumption}
	\begin{center}
	\setlength\tabcolsep{2pt}
	\setlength{\belowcaptionskip}{0pt}
    \renewcommand\arraystretch{1.0}
	\scriptsize
 	\vspace{-2.0em}
	\resizebox{\linewidth}{!}{
		\begin{tabular}{c|ccc}
			\toprule
			 & Pseudorange processing &Carrier phase processing& Total\\
			\midrule
	R-EDM   & 1.59 & 3.38 & 4.97  \\
	I-EDM   & 0.85  & 1.64  &2.49 \\
             \specialrule{0.08em}{1pt}{1pt}
		\end{tabular}
		\vspace{-2em}
		}
	\end{center}
\end{table}
\section{Conclusion}
The utilization of the GNSS/INS/Vision integration system provides a more comprehensive and robust positioning capability. However, multipath and cycle slips brought on by obstructions deteriorate GNSS location performance in urban areas, which further impairs multi-sensor integration positioning. For GNSS pseudorange and carrier phase measurements, this work develops an innovation-based cycle slip, multipath estimation, detection, and mitigation (I-EDM) technique. The cluster analysis method is utilized to extract the innovations from GNSS measurements, which are subsequently employed to identify cycle slips and multipath.

The experimental results show that the proposed strategy effectively reduces the impact of outliers within observations in urban test scenarios. Our proposed method significantly enhances the positioning accuracy when compared to the signal-to-noise ratio (SNR) model, $1/{sin\theta}^2$ satellite elevation angle model, and the hybrid model.  Moreover, a detailed comparison between the performance of residual-based and innovation-based EDM methods is conducted. Compared with the residual-based EDM method, the innovation-based EDM method obtains a decrease in average error rates of cycle detection by 57.1\%, a 21.6\% improvement in positioning accuracy, and a 49.9\% reduction in running time consumption. These results show the significant improvements achieved by the proposed innovation-based EDM method over the residual-based approach.

\section*{Acknowledgments}
Upon reasonable request to the corresponding author, the experimental data used in this research is available. 
This work is sponsored in part by The National Key Research and Development Program of China (2021YFB2501100), the National Natural Science Foundation of China (Major Program, Grant No. 42192533), and the Fellowship of China National Postdoctoral Program for Innovative
Talents (Grant No. BX20200251).




 
%


\bibliographystyle{IEEEtran}
\bibliography{ref}


 

\begin{IEEEbiography}[{\includegraphics[width=1in,height=1.25in,clip,keepaspectratio]{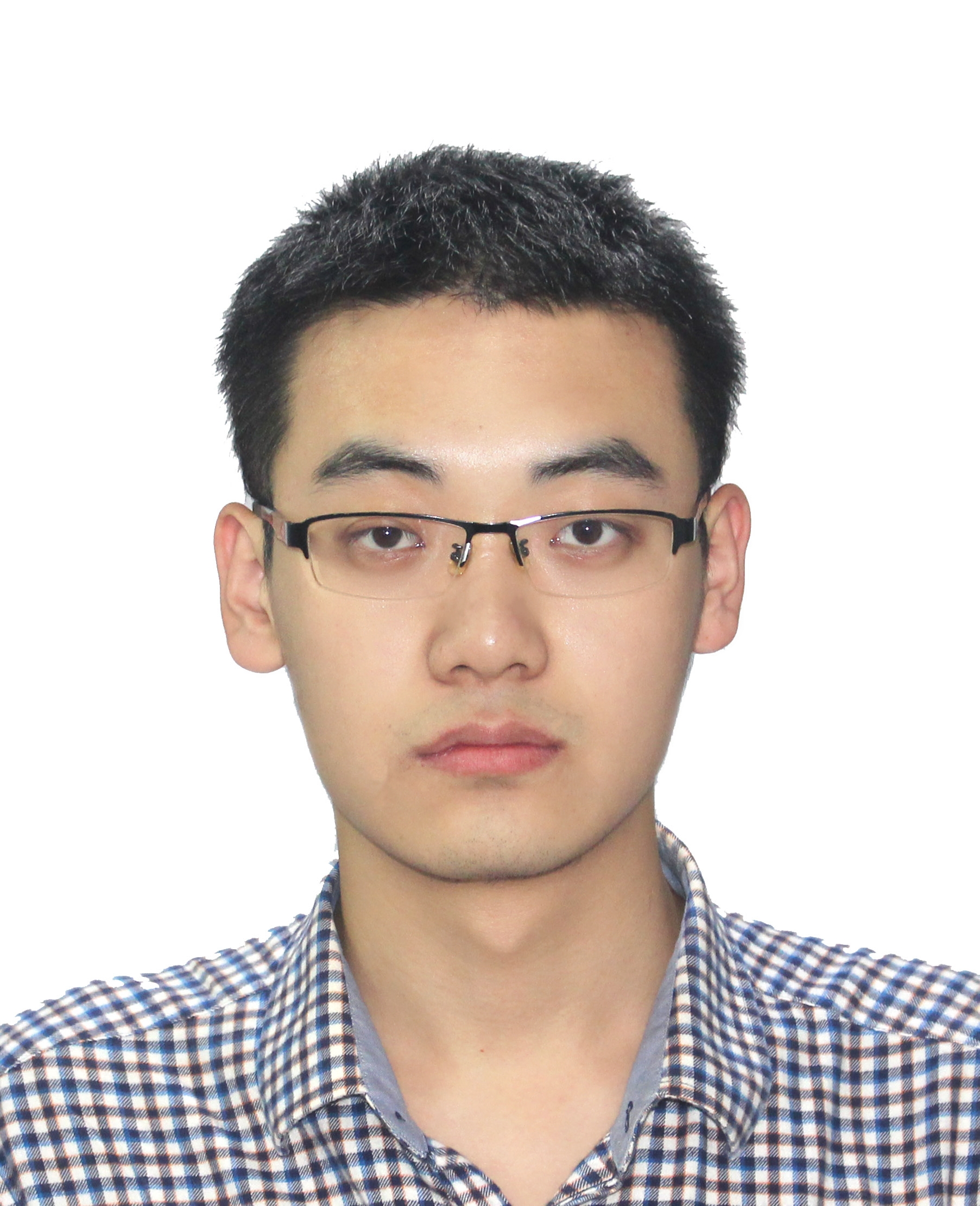}}]{Bo Xu} received the B.E. degree in the School of Land Science and Technology from China University of Geosciences, Beijing, China, in 2014 and 2018, and the M.E. degree in the School of Geodesy and Geomatics from Wuhan University, Wuhan, China, in 2018 and 2021, respectively, where, he is currently pursuing the Ph.D degree with the School of Geodesy and Geomatics. His research interests include GNSS precise positioning, visual SLAM, visual inertial odometry (VIO) and multi-sensor fusion algorithm.
\end{IEEEbiography}

\begin{IEEEbiography}[{\includegraphics[width=1in,height=1.25in,clip,keepaspectratio]{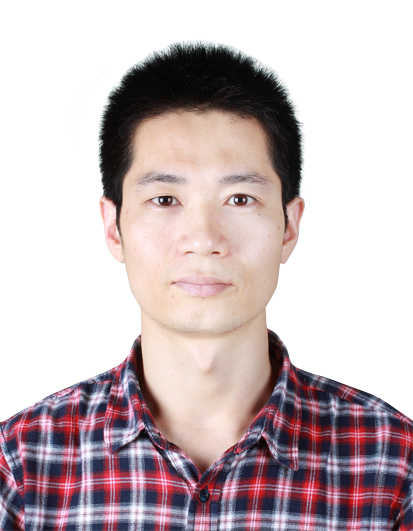}}]{Shoujian Zhang} received the B.Eng. and Ph.D.
degrees in the School of Geodesy and Geomatics from Wuhan University, Wuhan, China, in 2004 and 2009, respectively. He is an associate professor at the school of geodesy and geomatics from Wuhan University. His research interests are GNSS data processing,  multiple sensor  data fusion algorithms and applications.

\end{IEEEbiography}

\begin{IEEEbiography}[{\includegraphics[width=1in,height=1.25in,clip,keepaspectratio]{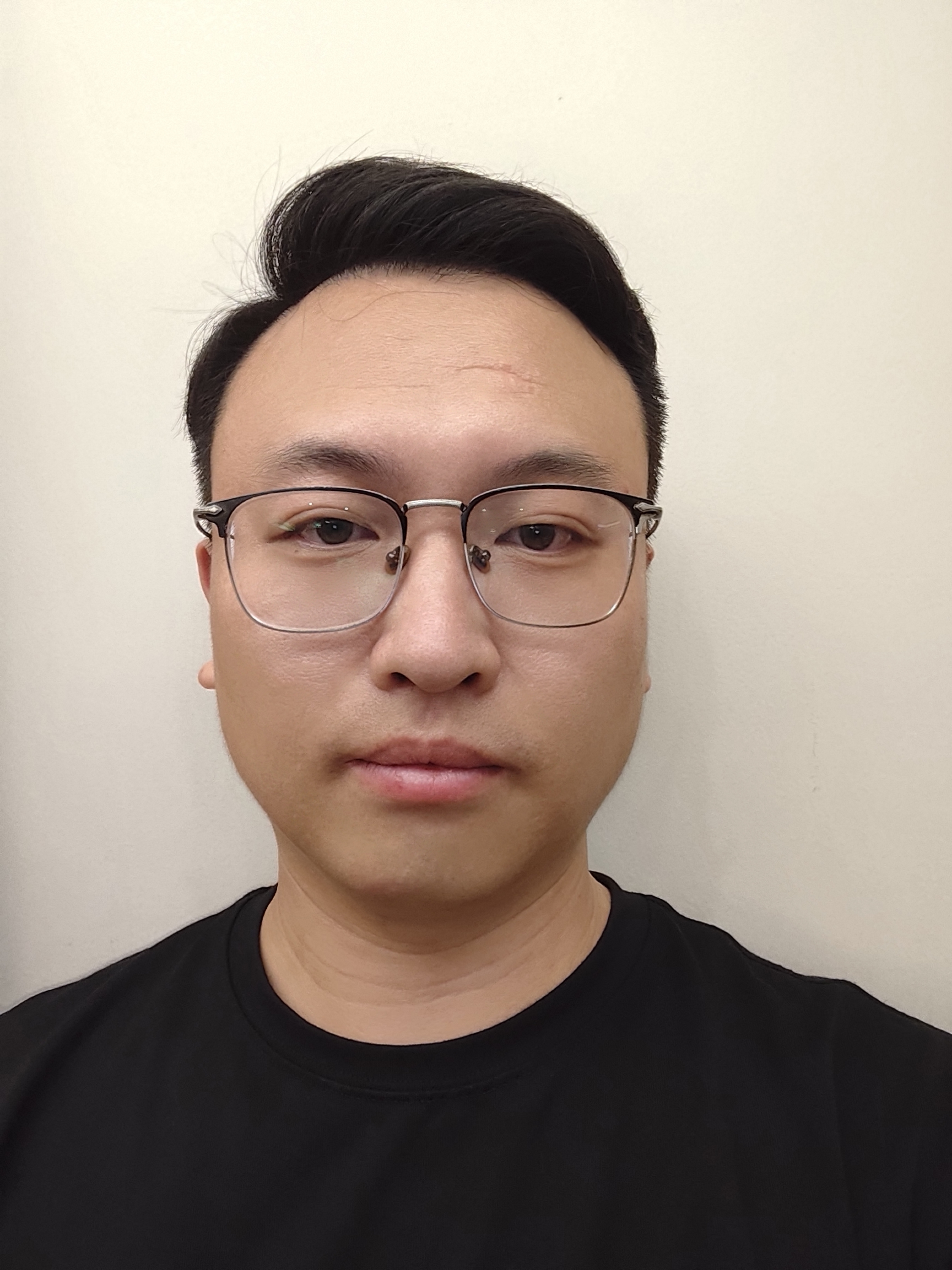}}]{Jingrong Wang}  received the B.E. degree in the School of Land Science and Technology from China University of Geosciences, Beijing, China, in 2014 and 2018, and the M.E. degree in  the State Key Laboratory of Information Engineering in Surveying, Mapping and Remote Sensing (LIESMARS) from Wuhan University, Wuhan, China, in 2018 and 2020, respectively, where, he is currently pursuing the Ph.D degree with the GNSS Research Center. His research interests include GNSS precise positioning, visual inertial odometry (VIO) and multi-sensor fusion algorithm. 
\end{IEEEbiography}

\begin{IEEEbiography}[{\includegraphics[width=1in,height=1.25in,clip,keepaspectratio]{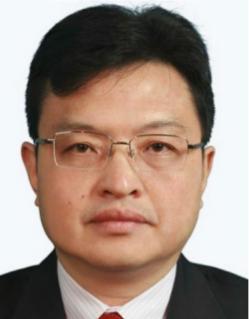}}]{Jiancheng Li} obtained his PhD degree from Wuhan University in 1993 and is a professor in geodesy at Wuhan University, Wuhan, China. He was honored as an academic member of the Chinese Academy of Engineering in 2011. His main research interests are the earth gravity field and its engineering applications.
\end{IEEEbiography}
\vfill

\end{document}